\documentclass[]{aa}
\usepackage{psfig}

\begin{document}

\title{On the global structure of distant galactic disks}
\author{V. P. Reshetnikov\inst{1,}\inst{2,}\inst{3}, R.-J. Dettmar\inst{2},
\and
 F. Combes\inst{4}}

\offprints{V.P. Reshetnikov resh@astro.spbu.ru }

\institute{Astronomical Institute of St.\,Petersburg State 
University, 198504 St.\,Petersburg, Russia 
\and 
Astronomisches Institut der Ruhr-Universit\"at Bochum, 
Universit\"atsstr. 150 NA7, 44780 Bochum, Germany
\and
Isaac Newton Institute of Chile, St.\,Petersburg Branch 
\and
LERMA, Observatoire de Paris, 61 Av. de l'Observatoire, 75014 Paris, France}

\date{Received 11 July 2002; accepted 3 December 2002}

\titlerunning{Distant disk galaxies}

\authorrunning{Reshetnikov, Dettmar, \& Combes}

\abstract{
Radial and vertical profiles are determined for 
a sample of 34 edge-on disk galaxies in the HDFs, selected for their apparent
diameter larger than 1.3\arcsec and their unperturbed morphology.
The thickness and flatness of their galactic disks are determined and discussed
with regard to evolution with redshift.
We find that sub-$L^*$ spiral galaxies with $z \sim 1$ have a relative
thickness or flatness (characterized by $h_z/h$ the scaleheight to scalelength 
ratio) globally similar to those in the local Universe.
A slight trend is however apparent, with the $h_z/h$ flatness 
ratio larger by a factor of $\sim1.5$ in distant galaxies if compared to 
local samples. In absolute value, the disks are smaller than in present-day 
galaxies. About half of the $z \sim 1$ spiral disks show a non-exponential surface
brightness distribution.
}

\maketitle
\section{Introduction}
A very fundamental characteristic of galactic disks is the flatness, defined
by the ratio between exponential scalelengths in the 
vertical and radial direction  $h_z/h$. 
The determination of the thickness of stellar disks is 
a difficult enterprise and has been determined mostly for edge-on
galaxies, although attempts have been made in other cases, too
(Ma et al. 1998). Typical values are found to be between 0.15 and 1
(van der Kruit \& Searle 1981, 1982; de Grijs 1998) with a systematic trend 
for galaxies to become thinner from S0 to Sc if a more complex 
multicomponent structure (thick/thin disk) is not taken into account.
Stellar velocity dispersion measurements show that the dispersion 
tends to grow with total luminosity, and vary radially as
the square root of surface density, so as to support
a constant scaleheight (Bottema 1993).
Recently Kregel et al. (2002) have shown that the flattening of disks
increases with the amplitude of their maximum rotation, and their total 
HI mass.

This flatness is an important clue to the
formation and evolution history
of galaxies. Disk stars are thought to be formed out of
a very thin gaseous disk. The observed structure then thickens due to
scattering by local fluctuations of the gravitational field as caused, e.g.,
by giant molecular clouds or  spiral structure in the disk.
The velocity dispersion of the stars increases in both the radial
and vertical direction by this diffusion through 
phase space (Wielen 1977, Binney \& Lacey 1988).
The disk may thicken more efficiently  through interactions 
with small companions or minor mergers,
as suggested by numerical simulations
(Quinn et al. 1993, Walker et al, 1996, Velazquez \& White 1999).
The thickening could amount to a factor of 1.5--2 and can be a constraint
on the frequency of mergers if disks are observed today to be too thin 
(e.g. Toth \& Ostriker 1992). 
The existence of a thick stellar disk in our own Galaxy has
been attributed to an old merger event (Robin et al 1996,
Dalcanton \& Bernstein 2002).
However, other phenomena have to be
taken into account since the thin disk could also be maintained
through gas infall with subsequent star formation.
Many observational facts strongly suggest a high gas infall rate
(e.g. Toomre, 1990, Sancisi et al. 1990, Jiang \& Binney 1999). Dynamics of
galaxy disks, bar and spiral reformation constrain this rate
such that galaxies may double their mass in less than 10 Gyrs (Bournaud \& 
Combes 2002, Block et al 2002).

The influence of tidal interaction on the flatness 
of galaxy disks  has been addressed by Reshetnikov et al. (1993)
and Reshetnikov \& Combes (1997) who found in interacting
galaxies  the ratio $h_z/h$ to be a factor of two  higher, due to a disk thicker
in absolute values, but also shorter in radial dimensions.
Schwarzkopf \& Dettmar (2000, 2001) confirmed this trend
on a larger sample, but essentially because of a larger
absolute thickness of interacting disks. They also
noted that tidally interacting galaxies are more
frequently warped and that the thickening was more noticeable 
in the outer parts. 

All these studies suggest that the relative flatness of a galaxy disk
can vary by both effects during its evolution since the majority of 
galaxies have experienced interactions in the past.  
It appears that galaxies were dynamically ``hotter'' in the past (Abraham
et al. 1999), which is seen in a smaller fraction of barred
galaxies, and this could also have some consequences for the
flattening. As the frequency
of interactions/mergers strongly  increases with redshift
(e.g. Le F\`evre et al 2000), it is interesting to trace the evolution 
of the disk flatness with time by studying photometrically distant
galaxies with high spatial resolution. This is now possible in the
Hubble Deep Fields HDF-N and HDF-S and we present below
the results of such a study. Section 2 describes our sample
and Sect. 3 gives the  derived general characteristics of disks in terms
of radial and vertical profiles. The last section discusses the
main results on the evolution of the flatness for  stellar disks.

Throughout the paper, we adopt a flat cosmology with $\Omega_0=1$ and
$H_0 = 70$ km~s$^{-1}$ Mpc$^{-1}$. (For a model with $\Omega_m=1/3$,
$\Omega_{\Lambda}=2/3$ and $\Omega_m+\Omega_{\Lambda}=1$, linear sizes
at $z=0.7-1$ will be $\approx$25\% larger and absolute magnitudes
are brigther by $\approx$0\fm45.) If not differently specified,
all magnitudes are expressed in the AB system (Oke 1974).

\section{The sample}

To study the structural parameters of distant edge-on spiral galaxies,
we consider a sample of galaxies presented in Paper I (Reshetnikov et al. 2002).
This sample includes 45 galaxies with  a diameter larger than 1\farcs3 
selected in the Hubble deep fields north and south. 
By applying a $V/V_{\rm max}$ completeness test we derived that
the  sample is statistically complete -- $V/V_{\rm max}$=0.50$\pm$0.04 
(Paper I). From the total sample of 45 distant galaxies we rejected 
objects with 
strong overlaps or interactions (e.g. n1031, SB-WF-2353-1914/2340-1898) 
and objects with unusual morphology (for instance, the "tailed" galaxies 
n16, n632, n1027). Also, we excluded the galaxies n749 and SB-WF-3329-3173 
since they seem to be deviating from the edge-on orientation. The final 
sample consists of 34 edge-on galaxies suitable for further analysis. 

The mean isophotal diameter of the galaxies within $\mu(I_{814}$)=26.0 is
2\farcs3 (14 kpc at $z=0.9)$, 
the mean minor axis is 0\farcs7. The latter is about 5 times larger than
the FWHM of the PSF and therefore the galaxies are sufficiently resolved
to study the vertical surface brightness distribution.

The general characteristics of the sample galaxies are presented in Table 1.
Figure~8 (see Appendix) gives contour maps and radial major axis profiles 
of the objects.

\begin{table}
\caption{Edge-on galaxies in the HDF-N and HDF-S}
\begin{center}
\begin{tabular}{|c|c|c|c|c|}
\hline
ID &    $I_{814}$& $b/a$  & $z$ &  Type  \\
\hline                   
   15 &  24.70  &  0.28  &0.45 &   Scd  \\
   55 &  23.32  &  0.20  &0.18 &   Irr  \\
   120&  24.27  &  0.24  &1.60 &   Irr  \\
   153&  24.94  &  0.35  &0.73 &   Irr  \\
   273&  22.47  &  0.45  &0.905&   Ell  \\
   304&  25.60  &  0.35  &1.78 &   Scd  \\
   450&  24.25  &  0.30  &0.70 &   Irr  \\
   476&  22.77  &  0.39  & 0.421&   Irr \\
   506&  24.47  &  0.25  & 0.751&   Irr \\
   534&  23.82  &  0.25  & 0.95 &   Sbc \\
   671&  23.93  &  0.36  & 0.681&   Irr \\
   716&  22.61  &  0.30  & 0.944&   Sbc \\
   727&  23.93  &  0.33  & 0.904&   Scd \\
   733&  25.04  &  0.25  & 0.94 &   Irr \\
   774&  21.22  &  0.28  & 0.485&   Sbc \\
   805&  25.33  &  0.30  & 0.66 &   Irr \\
   817&  22.19  &  0.40  & 0.485&   Sbc \\
   886&  25.25  &  0.30  & 0.97 &   Irr \\
   888&  23.70  &  0.30  & 0.559&   Irr \\
   898&  25.89  &  0.20  & 0.94 &   Irr \\
   899&  22.94  &  0.20  & 0.564&   Scd \\
   938&  23.03  &  0.27  & 0.557&   Scd \\
   979&  23.45  &  0.27  & 0.517&   Scd \\
 S0564&   25.27  & 0.35  & 1.02 &   Irr \\
 S0566&   25.67  & 0.30  & 2.16 &   Scd \\
 S0578&   23.16  & 0.22  & 0.98 &   Scd \\
 S0747&   25.31  & 0.24  & 0.98 &   Irr \\
 S1085&   25.51  & 0.29  & 1.13 &   Irr \\
 S1404&   23.09  & 0.20  & 0.50 &   Sbc \\
 S2661&   24.10  & 0.30  & 0.92 &   Irr \\
 S2691&   25.05  & 0.35  & 0.46 &   Irr \\
 S3053&   26.07  & 0.30  & 1.05 &   Scd \\
 S3458&   23.40  & 0.40  & 0.69 &   Scd \\
 S3685&   26.63  & 0.35  & 2.23 &   SB1 \\
\hline
\end{tabular}
\end{center}

The columns are: object identification (Paper I); $I_{814}$ magnitude in the AB 
photometric system according to Fern\'{a}ndez-Soto et al. (1999)
(aperture magnitude corrected for the effect of neighboring objects
and the flux lost in pixels outside the aperture); 
apparent axial ratio (from ellipse-fitting of the outer contours);
redshift (spectroscopic redshifts are with three numbers, photometric redshifts
obtained as the mean of redshifts presented by Fern\'{a}ndez-Soto et al. 1999
and by Fontana et al. 2000); best-fit spectral type of galaxy 
(Fern\'{a}ndez-Soto et al. 1999).
\end{table}

\section{General characteristics of edge-on disks}

As judged from to the spectral energy distributions (Fern\'{a}ndez-Soto 
et al. 1999), late-type galaxies (Sbc/Irr) predominate in our sample (Table 1). 
The mean redshift of the sample and the median of the distribution are the 
same, $z \approx 0.9$. The mean rest-frame absolute magnitude of the galaxies 
is relatively low -- $M(B) \approx -18{^{\rm m}}$ (Paper I). 
This is not unexpected since the Hubble deep fields have a very small 
coverage of only $\sim$2 Mpc (comoving) at $z \sim 1$ each. At $z < 1$, 
the expected number of bright $L^*$ galaxies in the two fields is 
only $\sim$20-60 (Ferguson et al. 2000). Therefore, assuming random 
orientation of galaxies, we can expect to find a few highly inclined bright 
spirals at $z < 1$. 

Two of our edge-on galaxies are possible sources of sub-mm (SCUBA)
radiation (Serjeant et al. 2002). These are the objects 304 and 774 in
Table 1. 

The mean observed axial ratio of the sample objects is 0.3 (Paper I, Table 1), 
a value exceeding the $b/a$ ratio for local edge-on late-type galaxies. 
Possible reasons are a large intrinsic thickness of distant objects 
and also the influence of the point-spread function (PSF) on the apparent 
thickness of the small ($\sim 2\arcsec$) HDF galaxies. 

The largest uncertaintiy arises from  the range of galaxy inclinations in
our sample since we do not know the true intrinsic flatnesses
 of the sample objects and we also  
cannot use a standard approach, e.g., on the basis of the Hubble formula
or from dust lane geometry.
We selected our sample on the basis of careful visual inspection of
the HDF frames. Secondary inclination indicators (asymmetry of
the minor axis profiles, apparent shift of the nucleus location from the
symmetry plane) are in the same range as in local samples of highly 
inclined spirals (e.g. de Grijs \& van der Kruit 1996, Schwarzkopf \&
Dettmar 2000). Also, comparison with results of realistic numerical
models for inclined spirals (e.g. Byun et al. 1994) indicates that most 
of our sample objects are within 5$^{\rm o}$ of edge-on orientation.
Such a moderate deviation from edge-on orientation does not significantly 
change the slope of the vertical surface brightness distribution
(van der Kruit \& Searle 1981, Barteldrees \& Dettmar 1994).

\subsection{Optical colors}

\begin{figure}
\centerline{\psfig{file=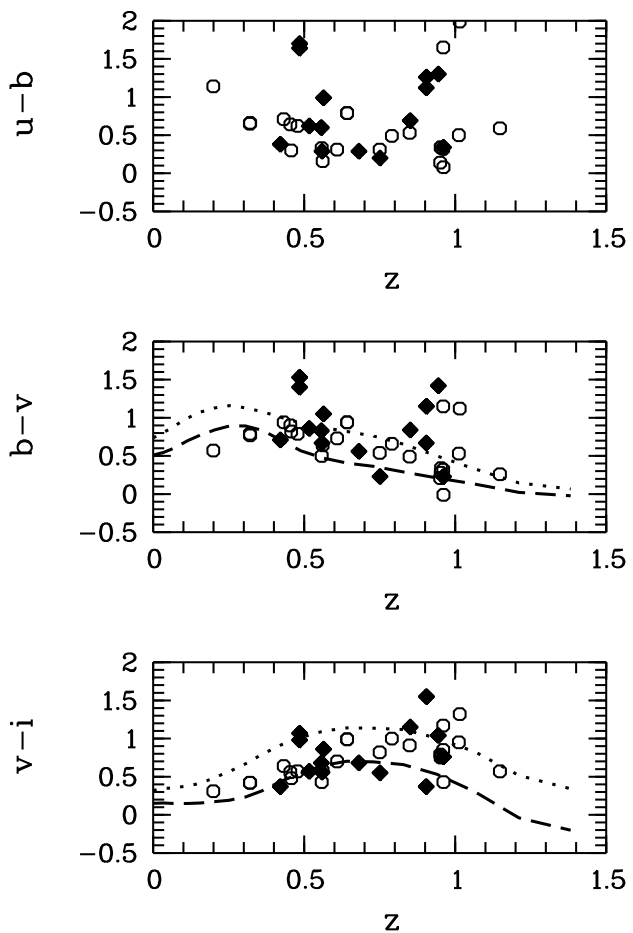,width=8.7cm,angle=-90,clip=}}
\caption{Observed color -- spectroscopic redshift dependence for the HDF 
edge-on galaxies (solid rhombs) and for face-on spirals (open circles).
The $u-b$, $b-v$, and $v-i$ corresponding to $U_{300}-B_{450}$, $B_{450}-V_{606}$, and
$V_{606}-I_{814}$ colors (Williams et al. 1996).
The lines show predicted color evolution as a function of redshift for
two models of disk galaxy formation -- the accretion (dashed line) and
the collapse (dotted line) models (Westera et al. 2002).}
\label{col}
\end{figure}

Figure~\ref{col} compares the observed colors of the 14 sample objects
with spectroscopic redshifts
and of nearly face-on spirals (randomly selected from the HDF-N) 
as a function of redshift. Both subsamples demonstrate the same trend of
colors with $z$ with a somewhat larger scatter for the edge-on disks.
In the redshift interval $z=0.4-1$ the edge-on disks show a hint of reddening
in comparison with the face-on spirals: $E_{u-b}=+0.25 \pm 0.18$ and
$E_{b-v}=+0.23 \pm 0.13$. These estimates correspond to inclination
corrections for local spiral galaxies (e.g. de Vaucouleurs et al. 1991).

The predictions of apparent color evolution for two models of galaxy
formation are shown in Fig.~\ref{col} according to Westera et al. (2002).
The first model is characterized by a slowly growing dark halo with a
continuous infall of gas and dark matter (accretion model, dashed line). 
The second one is a classical collapse model in a dark matter halo
(collapse model, dotted line). Obviously, both models provide a
satisfactory description of general observational trends of colors with
redshift. 

\subsection{Major axis profiles}

To investigate the global structure of the sample objects we adopted
as a fitting model 
the simple standard double exponential disk:
$L(r,z) = L_0$\,exp($-\frac{r}{h}-\frac{\mid z \mid}{h_z}$), 
where $h$ and $h_z$ are the scalelength and scaleheight, respectively.
This simple model gives a reasonable description of the structure
for most local disks, even in edge-on orientation 
(e.g. Misiriotis et al. 2000). Comparison with other vertical
models for disks (for instance, with ``sech$^2(z/z_0)$" or 
``sech$(z/z_0')$" distributions) can be done via 
$h_z = z_0/2 = z_0'/\sqrt{2}$. 

\subsubsection{Exponential profiles}

Figure~8 presents the major axis photometric
profiles of all galaxies in the $I_{814}$ passband. The profiles
show a large diversity in their shapes. Trying to fit the surface
brighness distributions by an exponential law, we find that
only 19 or 56\%$\pm$7\% of the disks are reasonably well described 
by  exponential-like profiles.

The parameters for the exponential disks are presented in Table 2.
Typical errors for the  scalelength values are $\approx$ 10\%. The errors of
scaleheights are presented in  Table 2 and they are not just the formal
errors of the least-squares fit, rather they  were estimated by comparing
results from four independent fits (see Sect. 3.3).
To check the reliability of the exponential model to describe
the global photometric structure of the studied galaxies, we have calculated
total luminosities as $L_{\rm exp }=2 \pi I_0 h^2 b/a$,
where $I_0$=dex(-0.4$\mu_0^I$) (Table 2) and $b/a$ is the apparent
axial ratio (Table 1). We found that the mean difference
of observed and model magnitudes is $\langle I_{814} - I_{\rm exp } \rangle$= +0.31$\pm$0.12. 
This small difference can be attributed to the infinite extrapolation
for the model magnitudes.

The mean observational characteristics of the exponential disks are
$\langle \mu_0(I_{814}) \rangle = (22.99 \pm 1.16) {^{\rm mag}}/\sq\arcsec$, 
$\langle h \rangle = 0\farcs43 \pm 0\farcs11$, $\langle z \rangle = 1.0 \pm 0.5$. 
The observed values of $\mu_0(I_{814}$ (uncorrected for any inclination
effects) have been converted to a rest-frame $B$ by applying the cosmological
dimming term and a $k$-correction color term for Irr galaxies
(Lilly et al. 1998, Lilly et al. 1995, Paper I). 
The resulting mean characteristics of the central surface brightness
and of the scale length (after a small correction for the PSF -- see Sect.3.3)
are $\langle \mu_0(B) \rangle = (20.7 \pm 1.3) {^{\rm mag}}/\sq\arcsec$,
$\langle h \rangle = 2.3 \pm 0.7$ kpc. For the disks with
$z \leq 1$ (14 objects), the central surface brightness
is (21.05$\pm1.0){^{\rm mag}}/\sq\arcsec$. 
Therefore, distant edge-on disks demonstrate
somewhat brighter central surface brightnesses in comparison
to nearby "Freeman" disks for which $\mu_0(B)=21.5{^{\rm mag}}/\sq\arcsec$ 
in the AB
system (Freeman 1970). However, this difference is negligible within
the quoted scatter. Note also that our derived value of the mean central
surface brightness is in good agreement with  results for
local samples of edge-on galaxies: $\langle \mu_0(B) \rangle = 
(21.0 \pm 0.6){^{\rm mag}}/\sq\arcsec$
(28 members of interacting systems, Reshetnikov \& Combes 1997),
$\langle \mu_0(B) \rangle = (21.0 \pm 1.1){^{\rm mag}}/\sq\arcsec$ 
(45 normal galaxies, de Grijs 1998). 

As a first order approximation, let us assume that the galaxy
centers are optically thick, or close to that, and so the
central surface brightness does not depend strongly on the  
inclination of the disk.
This is a quite realistic approximation since at redshift $z = 0.9$ 
the filter $I_{814}$ corresponds approximately to filter $B$ 
(8140 \AA:(1+0.9)=4280 \AA) and the central regions of local spirals are not optically thin
in the blue spectral region. There are several lines of arguments
-- discussed in the following -- in favour of such a conclusion.

The only direct and model-independent way to study the transparency
of the centers of galaxies is the chance superposition of galaxies. 
In the innermost
few hundred parsecs of NGC~3314A (a sub-$L^*$ Sc spiral galaxy) even
the most transparent regions between dust lanes show an extinction of
$A_B \approx 7^m$ (Keel \& White 2001). The disk of NGC~3314A is inclined 
by $i=41^{\rm o}$, so this galaxy is far from edge-on orientation. 

The same conclusion can be derived from statistical arguments.
For instance, Giovanelli et al. (1995), using a sample of more than
1700 Sc galaxies with $I$-band photometry, found only a weak dependence 
of the central surface brightnesses of the disks on inclination: 
$\mu_0(I)$ brightens by about 0.3--0.4 mag between the face-on and edge-on
aspects. In the $B$ band the dependence must be even much weaker. 
Also, with regard to  distant galaxies, Lilly et al. (1998) 
did not find any significant correlation between $\mu_0(B)$
and disk axial ratio for spirals with $z = 0.2-1$.

Numerical simulations of the radiative transfer in dusty disks show
that even a small optical depth can significantly reduce the expected
projection effect on the central surface brightness, especially in
the $B$ band (Byun et al. 1994). 

Therefore, assuming that the central surface brightness 
shows only a weak (or negligible) dependence on the galaxy inclination 
($\Delta \mu_0(B)=(0-0.5)^{\rm mag}/\sq\arcsec$)
we can estimate an average ``face-on" absolute magnitude of the disks
as $\langle M(B) \rangle = \langle \mu_0(B) \rangle$ -- 
5\,log$\langle h \rangle$ -- 38.57 = --19\fm6...--19\fm1. 
(Let us note also that in edge-on orientation the presence of 
dust makes the scalelength appear larger than in the face-on case 
(e.g. Byun et al. 1994) and this can reduce the estimated
face-on luminosity.) The observational
value for edge-on disks is --18\fm1 or 1\fm0...1\fm5 fainter. 
The value of dimming (1\fm0...1\fm5) is consistent within 
uncertaintes with the value of global extinction for spirals 
(e.g. de Vaucouleurs et al. 1991, Tully et al. 1998).

Table 2 presents also the $h$ values in the $V_{606}$ passband.
 The mean scale length ratio is 
$\langle h(I_{814})/h(V_{606}) \rangle$ = 1.04$\pm$0.13.
(In the rest-frame this ratio resembles 
$h(B)/h(U)$.) Although we did not find any significant color dependence
as expected from the modeling by de Jong (1996) for these filters,
we cannot draw any firm conclusions due to the small-number
statistics. 

\begin{figure}
\centerline{\psfig{file=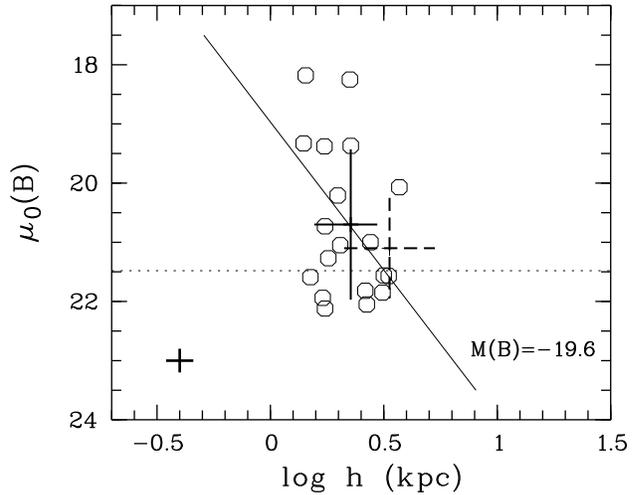,width=8.3cm,angle=-90,clip=}}
\caption{$\mu_0(B)$--$h$ plane for edge-on galaxies. 
 Typical error of the data is shown by a small cross at bottom left.
The dotted line shows the Freeman law, the solid line is the line of 
constant disk luminosity. The solid cross gives the mean characteristics 
of our edge-on disks and the dashed cross is the mean for the sample of
917 local late type spirals with $M(I) \geq -22^m$ from Byun (1992).}
\label{muh}
\end{figure}

In Fig.~\ref{muh} we consider the $\mu_0$--$h$ plane. 
The dashed cross in the figure shows
the mean parameters of 917 Sb--Sd spiral galaxies from Byun's
(1992) thesis (converted to $B_{AB}$). The sample of Byun is
angular diameter-limited (like ours) and contains galaxies with angular
diameter $d \geq 1\farcm7$. The mean absolute magnitude
of this sample of local galaxies is $\langle M(I) \rangle = -20\fm4$, or
$\langle M(B) \rangle \approx -18\fm7$ (adopting $B-V$=+1.7 for
the disks of galaxies, de Jong 1996). Therefore, in edge-on orientation
the local galaxies will show about the same 
luminosities as our sample galaxies. As one can see in the figure,
the mean characteristics of nearby and distant sub-$L^*$ disks 
are close (within quoted uncertaintes) but $z \sim 1$ galaxies
show a small displacement to higher values of surface brightness and 
smaller sizes. The shift to smaller values of $h$ can be even larger
due to the above mentioned effect of an apparent increase of scalelength
in edge-on orientation. 

\begin{table*}
\caption{Characteristics of galaxies with exponential-like surface brightness 
distribution}
\begin{center}
\begin{tabular}{|c|c|c|c|c|c|c|c|}
\hline
ID &     $\mu_0^I$ & $h^I$        & $\mu_0^V$ & $h^V$  &  $h^I_z/h^I$ 
& $h^I$ & $h^I_z$ $\pm$ $\sigma_{h_z}$  \\
   &               &   ($''$)     &           &  ($''$)            &
& (kpc) &      (kpc) \\                     
\hline                   
   15 &   22.94      & 0.38   & 23.27 & 0.46 &    0.18 & 1.70 &  0.31~~ 0.06 \\
   153&   23.17      & 0.29   & 23.69 & 0.28 &    0.21 & 1.50 &  0.32~~ 0.10 \\
   304&   23.70      & 0.32   & 24.05 & 0.35 &    0.27 & 1.73 &  0.47~~ 0.25 \\
   450&   23.34      & 0.48   & 23.79 & 0.48 &    0.33 & 2.62 &  0.86~~ 0.34 \\
   476&   21.67      & 0.40   & 22.22 & 0.46 &    0.27 & 1.74 &  0.47~~ 0.09 \\
   671&   22.54      & 0.39   & 23.25 & 0.37 &    0.23 & 2.03 &  0.47~~ 0.16 \\
   727&   22.28      & 0.36   & 22.50 & 0.34 &    0.18 & 1.98 &  0.36~~ 0.06 \\
   733&   23.76      & 0.59   & 24.54 & 0.70 &    0.17 & 3.32 &  0.56~~ 0.11 \\
   774&   20.42      & 0.31   & 21.57 & 0.32 &    0.19 & 1.40 &  0.27~~ 0.03 \\
   805&   23.56      & 0.34   & 24.03 & 0.35 &    0.14 & 1.74 &  0.24~~ 0.07 \\
   817&   20.46      & 0.47   & 21.55 & 0.47 &    0.4  & 2.26 &  0.90~~ 0.21 \\
   888&   22.52      & 0.37   & 23.03 & 0.35 &    0.24 & 1.80 &  0.43~~ 0.07 \\
   898&   24.24      & 0.47   &       &      &    0.17 & 2.66 &  0.45~~ 0.09 \\
 S0566&   24.00      & 0.42   & 24.43 & 0.52 &    0.18 & 2.24 &  0.40~~ 0.10 \\
 S0578&   22.36      & 0.64   &       &      &    0.25 & 3.70 &  0.93~~ 0.20 \\
 S0747&   23.85      & 0.57   & 24.36 & 0.56 &    0.13 & 3.16 &  0.41~~ 0.08 \\
 S1085&   23.69      & 0.51   & 24.00 & 0.41 &    0.10 & 2.76 &  0.28~~ 0.03 \\
 S3053&   24.38      & 0.57   & 25.4  & 0.7  &    0.11 & 3.12 &  0.34~~ 0.14 \\
 S3685&   24.00      & 0.28   & 23.86 & 0.24 &    0.22 & 1.43 &  0.31~~ 0.11 \\
\hline
\end{tabular}
\end{center}
The columns are: galaxy identification; central surface brightness
and exponential scale length in arcsec ($I_{814}$ passband); the
same for the $V_{606}$ filter; scale height to scale length ratio
corrected for the PSF; final values of scale length and scale height
in kpc.
\end{table*}

At least three objects (n15, n805, and n898) can be considered as
low surface brightness galaxies. These three galaxies show 
the rest-frame central surface brightness $\mu_0(B) \approx 22$ and
are intrinsically faint ($M(B) \geq -17{^{\rm m}}$). Adopting small or
neglible internal absorption for such galaxies (e.g. Tully et al. 1998)
and converting the observational values of $\mu_0(B)$ to face-on
orientation, we obtain $\mu^0_0(B) \approx 23.5-24{^{\rm mag}}/\sq\arcsec$,
 typical for local low surface brightness galaxies (e.g. UGC~7321 --
Matthews \& Wood 2001).

\subsubsection{Non-exponential profiles}

15 of 34 disks (44\%$\pm$5\%) show ``ridge"-like profiles, with several peaks
and  are often very asymmetric (see, e.g., n55, n506, or n886 in Fig.~8). 
Therefore,  roughly half  of $z \sim 1$ disks are not 
exponential. This fraction may be even larger due to preselection
of the sample (Sect.2). This strongly contrasts with the local Universe
and supports a late formation of disks (e.g. Mo et al. 1998, 
see van den Bergh et al. 2000 about galaxy morphology evolution with
redshift).

\begin{figure}[!ht]
\centerline{\psfig{file=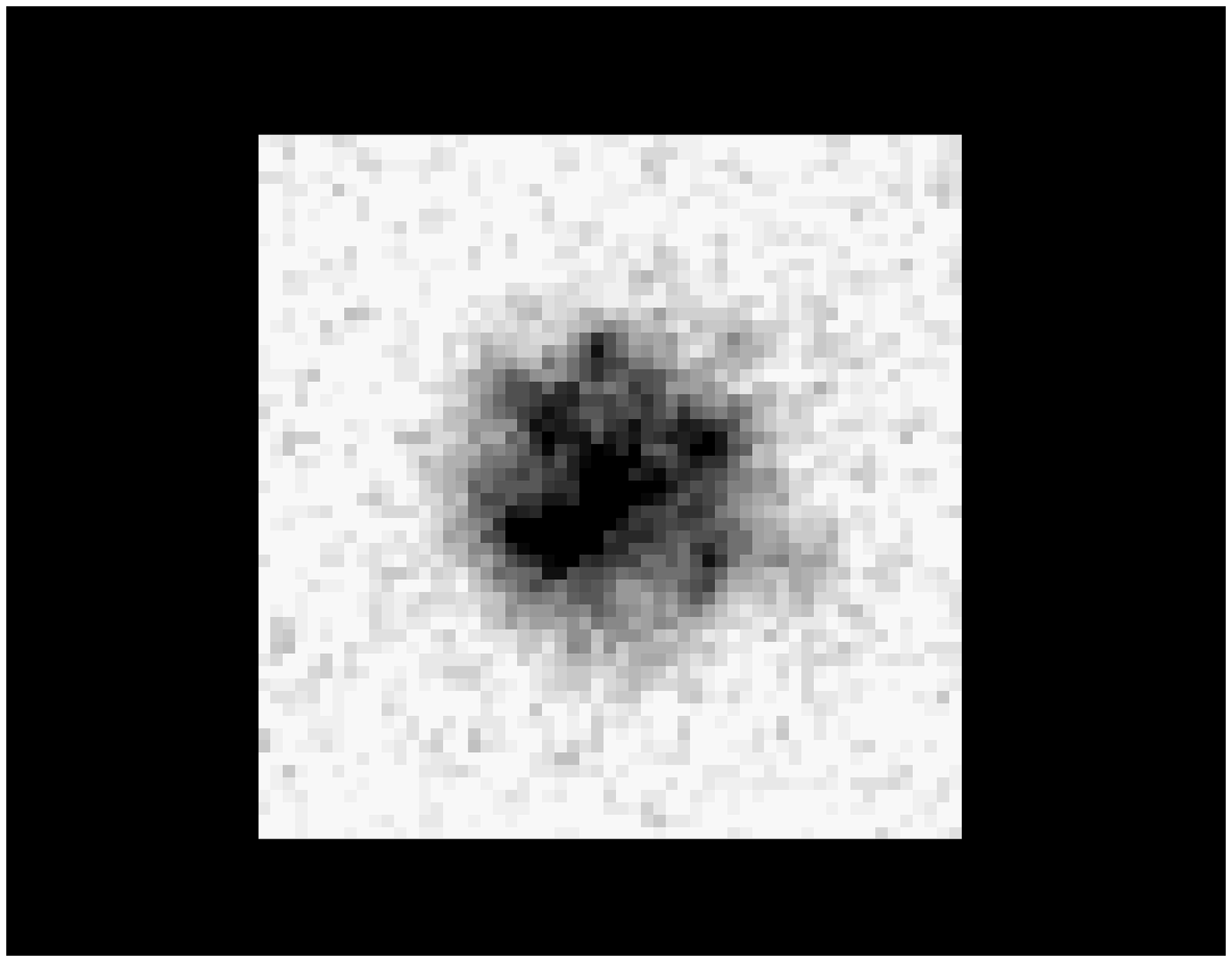,width=7.0cm,clip=}}
\caption{Reproduction of the HDF-N galaxy 866 (2.\arcsec3$\times$2.\arcsec3)
in the $I_{814}$ passband.}
\label{example}
\end{figure}

\begin{figure}[!ht]
\centerline{\psfig{file=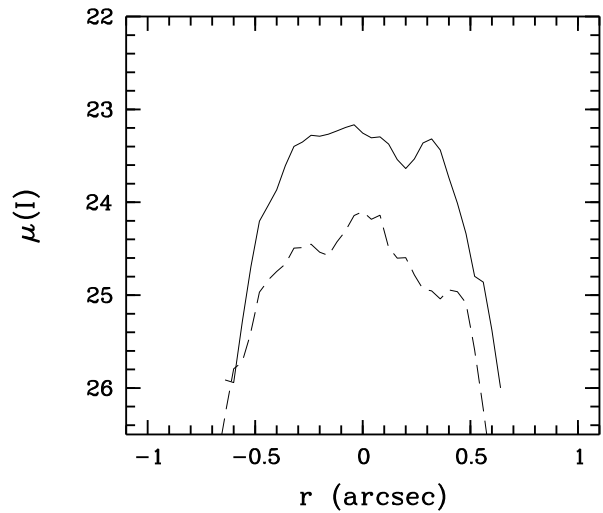,width=8.7cm,angle=-90,clip=}}
\caption{Photometric profiles of HDF-N 866. Solid line -- P.A.=125$^\circ$,
dashed line (the profile shifted down by 1$^{\rm m}$) -- P.A.=35$^\circ$.
The P.A. orientation is according to Fig.~\ref{example} (from top to the left).}
\label{excuts}
\end{figure}

To understand the origin of non-exponential brightness distributions,
we have looked for possible counterparts for such objects among
face-on galaxies in the deep fields. One of the best examples is
the HDF-N galaxy 866 (2-153.0 in Williams et al. 1996)
with the photometric redshift 0.96, 
$I_{814}=23.6$ and the spectral type Scd (Fern\'{a}ndez-Soto et al. 1999).
The observed characteristics of this galaxy are close to those
for the sample edge-on objects. This galaxy exhibits an irregular patchy
disk with a low-contrast nucleus and embedded bright knots (five at 
least -- see Fig.~\ref{example}) and no spiral structure has 
developed yet. Figure~\ref{excuts} shows two photometric profiles
of the galaxy which resemble the ``ridge-like'' profiles of many of our edge-on
disks mentioned above. The galaxy is very similar in optical
morphology to the peculiar galaxy H36490\_1221 discussed in van den Bergh
et al. (2001). Those authors suggest that such objects
represent  proto-late-type galaxies. Therefore, we can speculate
that about half of our sample objects are young disks still in the
formation process which will in their future evolution relax to an
exponential distribution. 

To check the frequency of non-exponential profiles among less distant
spirals we selected in two deep fields all bright ($I_{814} < 22^m$) and
non-edge-on galaxies of Sbc-Irr types. In total, we selected 17 galaxies
with mean redshift $\langle z \rangle \approx 0.5$ and absolute
magnitude $\langle M(B) \rangle \approx -20^m$. Almost one third
(5/17) of these objects show strongly non-exponential surface
brightness distributions (work in preparation). The rest of the objects
often demonstrate significant ($\sim 0\fm5$) deviations from
the exponent.
Therefore, non-exponential density distributions of galaxy disks are 
indeed very frequent at $z > 0$. 

\subsection{Minor axis profiles}

The typical scalelengths of our objects are $\approx 0\farcs3-0\farcs6$
(Table 2). Therefore, $h_z$ values are expected to be $\leq 0\farcs1$ or
below the PSF width of FWHM(PSF)=0\farcs14. 
It is important to note that $h_z$ values are 
characteristics of the surface brightness {\it gradients} and 
are not linear sizes of any structures. To determine
true scaleheights, we must study the effects of the HDF's PSF on the 
surface brightness distribution of edge-on galaxies.

We have created a grid of double exponential disk models with
various input parameters in the region $h=0\farcs3-0\farcs6$
and $h_z/h$ from 0.10 to 0.35. Then we prepared an empirical PSF
by averaging several stellar images from the $I_{814}$-band frames
of the HDF and convolved the model disks with the PSF. 
Comparing input and output disk characteristics, we determined
empirical corrections for the observed parameters. The
output parameters were estimated from the analysis of major and 
minor axes cuts of the model images. Finally, we found
a strong influence of the PSF on the vertical scale of a disk.
Also, we noted a small (within 10\%) overestimation of the
scale length values due to the PSF. 

\begin{figure}
\centerline{\psfig{file=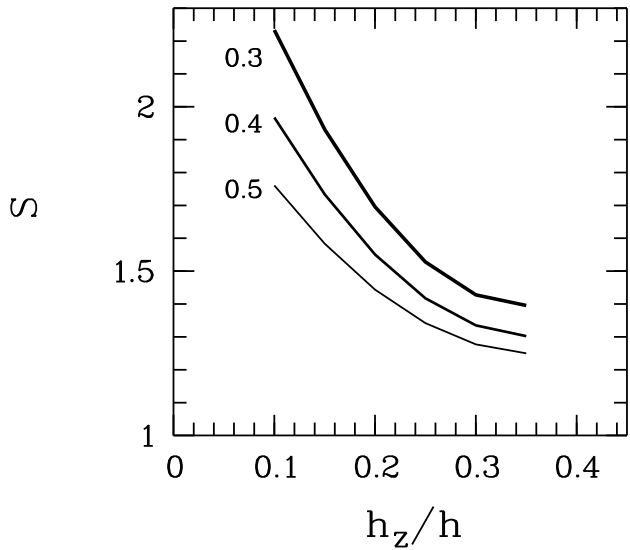,width=8.7cm,angle=-90,clip=}}
\caption{Dependence of the "stretching" factor S due to the PSF on the true
scaleheight to scalelength ratio $h_z/h$. The lines marked as 0.3, 0.4, 
0.5 correspond to models with $h$=0\farcs3, 0\farcs4, 
0\farcs5.}
\label{str}
\end{figure}

To present the influence of the PSF more clearly, we show in Fig.~\ref{str}
the dependence of vertical stretching due to the PSF on the
intrinsic flattening ($h_z/h$) of the model (the factor S gives the
ratio of observed and true values of flattening). 
It is evident that intrinsically small
($h \approx 0\farcs3$) and thin ($h_z/h \approx 0.1$) galaxies 
become more than 2 times thicker after application of the PSF.
Therefore, we cannot expect to find very thin galaxies among
small and distant HDF objects.

We excluded the central regions of the galaxies from the study in order
to avoid any possible contribution of the bulges. (Although we do not see
clear signs of bulges present in most galaxies -- see
Fig.~8). In order to explore the vertical
structure of 19 exponential objects we examined one-dimensional
cuts extracted at $\pm$(0.5--1)$h$ in the galactocentric distance, 
avoiding the faintest outer regions of the objects. 
To minimize the influence of dust, we excluded the 1--2 central pixels
of the cuts. In the analysis, we distinguished between the different
sides of the galaxy planes and obtained the mean
value of $h_z$ for each galaxy from 4 independent estimates. 
The mean observational (uncorrected for the PSF influence) scale height 
for the sample objects is 0\farcs13, the mean dispersion 
$\sigma_{h_z}=0\farcs03$. This value of dispersion (23\%) characterises
the combined effect of the intrinsic variations of vertical
scales and of errors in scale estimates. 

The results of the analysis 
(corrected for the PSF) are given in the last three columns of Table 2. 
The mean relative thickness of the galaxies reduces after the PSF
correction from $\langle h_z/h \rangle_{obs} = 0.31 \pm 0.09$ to 
$\langle h_z/h \rangle = 0.21 \pm 0.08$. 

\subsubsection{Vertical scale}

The mean value of the vertical scale for distant edge-on galaxies is
$\langle h_z \rangle$=0.46$\pm$0.21 kpc or
$\langle z_0 \rangle \approx0.9$ kpc. 
 The obtained mean value is unbiased since we work with a statistically
complete sample of galaxies with angular diameter larger than
1\farcs3 (Sect. 2, Paper I). These values are
typical for normal spiral disks at $z \approx 0$ 
(e.g. Reshetnikov \& Combes 1997, Schwarzkopf \& Dettmar 2000).
However, the direct comparison of local and distant samples 
must be carried out with  caution due to very different 
selection effects. For instance, distant galaxies are sub-$L^*$
and relatively compact while the local samples usually contain
brighter and larger disk galaxies (see Table 2 in Reshetnikov \& Combes 1997).

\begin{figure*}
\centerline{\psfig{file=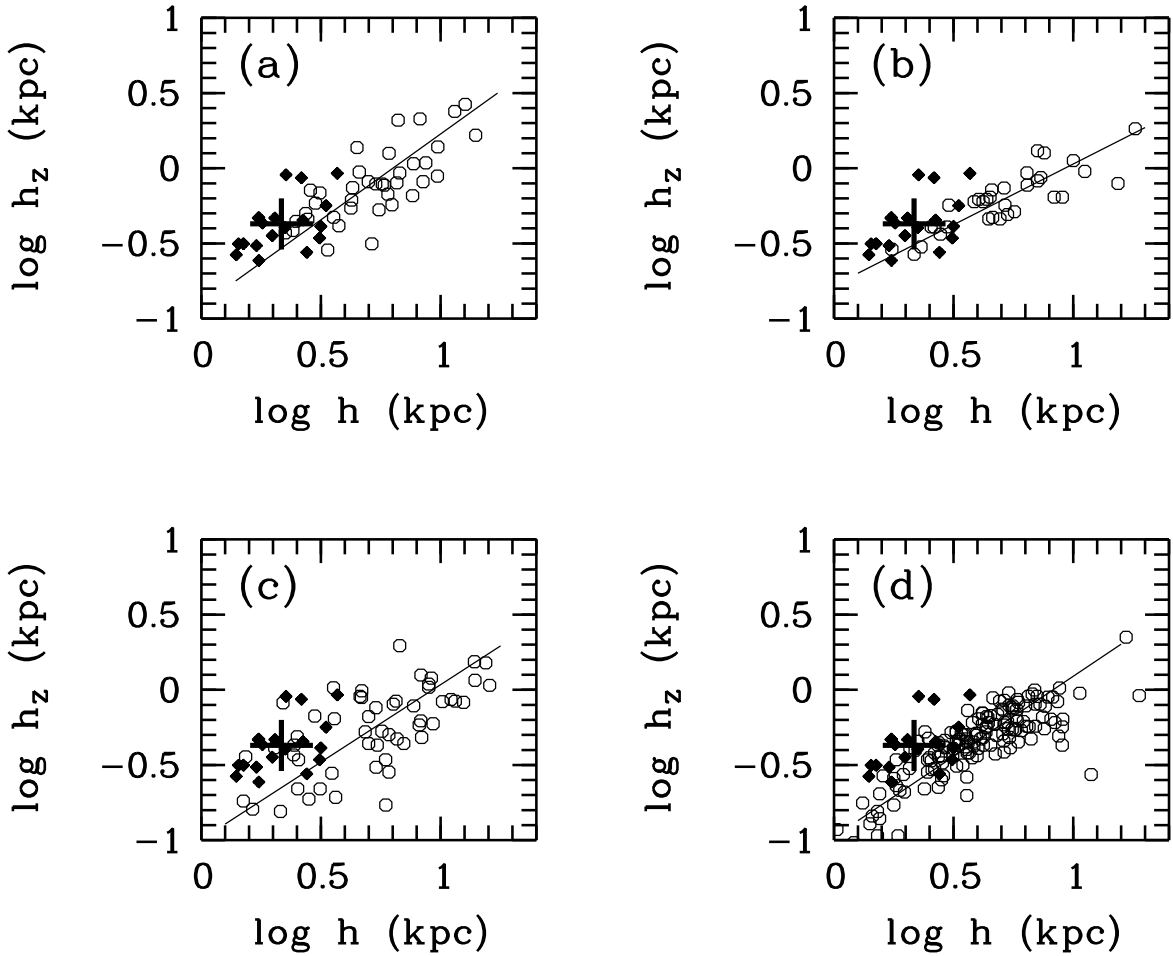,width=16.0cm,angle=-90,clip=}}
\caption{Scalelength -- scaleheight relation for local spirals
(open circles, (a) -- de Grijs 1998, (b) -- Kregel et al. 2002,
(c) -- Schwarzkopf \& Dettmar 2000, 
(d) -- Bizyaev \& Mitronova 2002) and for our distant galaxies (solid rhombs).
The solid lines show mean regressions for the nearby samples,
the cross represents the average characteristics of distant
galaxies and a $1\sigma$ scatter is indicated.}
\label{lenhei}
\end{figure*}

In order to avoid (at least partially) the selection effects,
we decided to compare our sample to local
galaxies of the same size.
Figure~\ref{lenhei}a shows the dependence of scaleheight
on scalelength for 38 edge-on spiral galaxies ($T > 0$) in the $I$
passband according to de Grijs (1998). He adopted an
exponential model for the observational brightness distribution 
and considered one-dimensional
photometric profiles so we can directly compare our results. 
The mean regression for
local galaxies predicts that a galaxy with $h=2.3$ kpc must have
$h_z=0.3$ kpc. Therefore distant galaxies are thicker by
a factor 1.5 on average.
(In the non-zero $\Lambda$ cosmology with $\Omega_m=1/3$, 
$\Omega_{\Lambda}=2/3$ the relative thickening stays the same.)

Figure~\ref{lenhei}b presents the results of a two-dimensional 
decomposition in the $I$ passband of 34 edge-on spirals from the
de Grijs (1998) sample recently presented by Kregel et al. (2002). 
The fitting model assumes a transparent disk (this is very
questionable for our sample objects) and includes the 
line-of-sight integration of the luminosity distribution. Comparison of two
methods shows that a one-dimensional method can overestimate the scalelengths
by about 20\% (Figure 3 in Kregel et al. 2002). 
However, a direct comparison of $h$ values shows that
$h$(de Grijs)/$h$(Kregel et al.)=0.93$\pm$0.20. 
A comparison of scaleheight values obtained by the two methods leads to
a ratio $h_z$(de Grijs)/$h_z$(Kregel et al.)=1.13$\pm$0.22.
Strictly speaking, both methods give a statistically insignificant
difference. The Kregel et al. (2002) analysis  results in a $\sim$1.4 
thickening of distant disks.

In Figure~\ref{lenhei}c we compare the Schwarzkopf \& Dettmar (2000)
sample of non-interacting galaxies (the data in the 
Johnson $R$ or Thuan \& Gunn $r$ filters). The fitting model
avoids the central dust lane in the line-of-sight integration. 
In this case the thickening factor increases to 1.9.

The last sample (Figure~\ref{lenhei}d) includes 153 edge-on
spiral galaxy with published photometric parameters in the $K_s$ band
(Bizyaev \& Mitronova 2002). The fitting model assumes a transparent
disk. This work gives about a 1.8-fold thickening of distant galaxies. 

\subsubsection{$h/h_z$ ratio}

The relative thickness -- $h/h_z$ -- gives more reliable 
information about the possible thickening. For our sample
we have $\langle h/h_z \rangle$=5.4$\pm$2.0, or,
excluding one object for being most likely not exactly an edge-on galaxy (n817),
5.55$\pm$1.9. Scaling this in terms of $z_0$, 
we obtain $\langle h/z_0 \rangle$=2.7--2.8, with a median value of
2.63. The disks of local late-type galaxies are distributed over a wider
range with the mean $h/z_0$ of $\approx 5$ (see Table 2 in 
Reshetnikov \& Combes 1997).  
Moreover,  disks of more late-type and gas-rich galaxies are thinner
(Reshetnikov \& Combes 1997, de Grijs 1998, 
Schwarzkopf \& Dettmar 2000). For Sc/Sd disks Schwarzkopf \& 
Dettmar (2000) give $h/z_0 \approx 8 \pm 2$. De Grijs (1998)
and Kregel et al. (2002) give somewhat smaller ratios of
3.7 and 4.3, respectively. The Bizyaev \& Mitronova (2002) sample
shows the mean ratio of $h/z_0 = 4.8$. 

In the previous discussion we neglected the possible dependence
of scalelength on the passband. It is known (e.g. Table 7 in de Grijs 1998)
that the $h$ value increases to the blue spectral region. The ratio
of scalelength values in the $B$ and $K$ passbands can reach
$\approx 1.5$, and  $h(B)/h(I)\approx 1.3$ (de Grijs 1998). 
This effect, however, would  only strengthen the thickness difference
in the samples of local and $z \sim 1$ galaxies. 

Therefore, $z \sim 1$ disks are definitely thicker
in comparison to local galaxies. The factor for the vertical
thickening ($\geq$1.5) is approximately the same as
for $z \approx 0$ interacting spiral galaxies ($\sim$1.5-2:
Reshetnikov \& Combes 1997, Schwarzkopf \& Dettmar 2000).

\begin{figure}
\centerline{\psfig{file=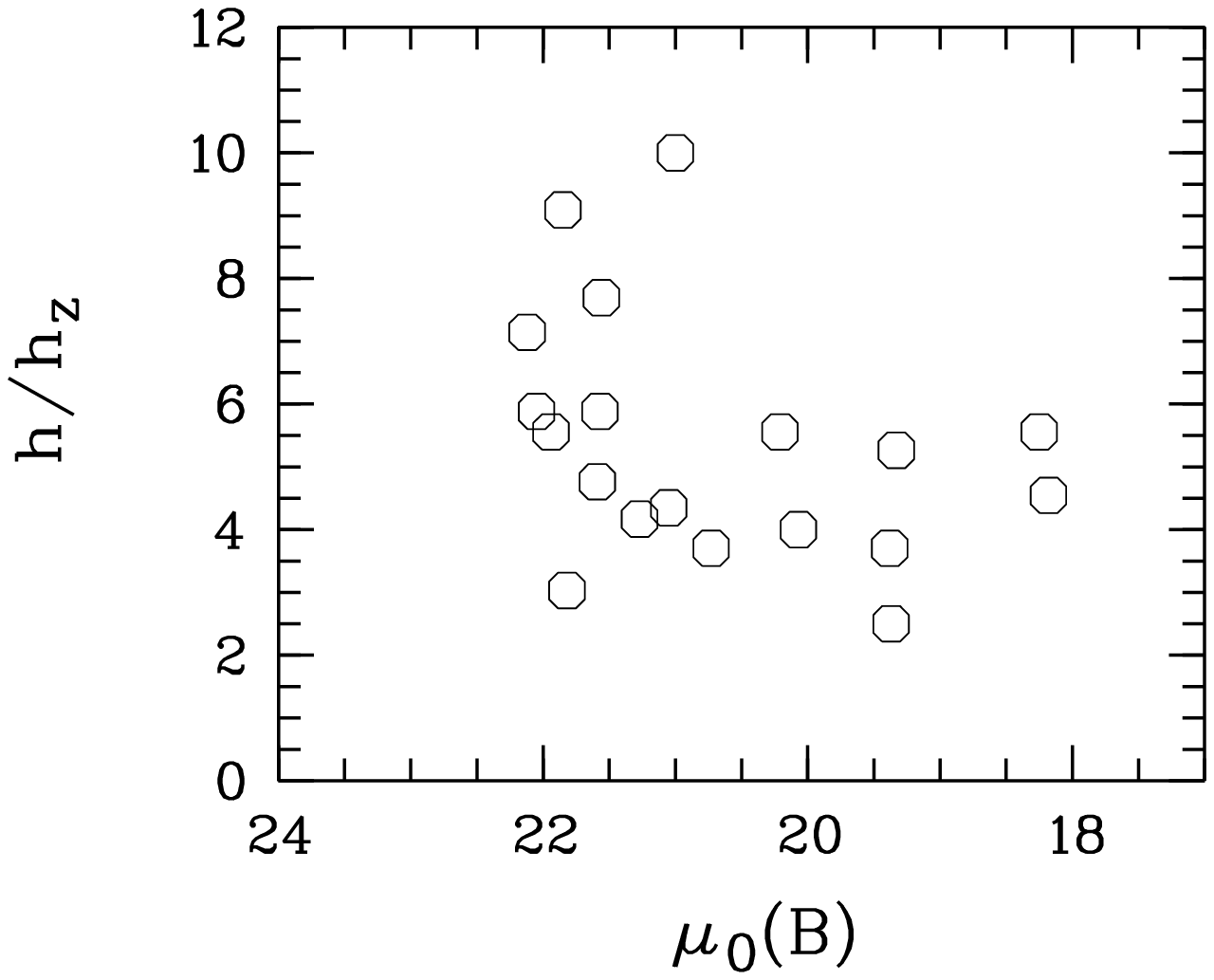,width=8.7cm,angle=-90,clip=}}
\caption{Dependence of the relative thickness of distant galaxies
on their rest-frame central surface brightness.}
\label{thickmu}
\end{figure}

The dependence of $h/h_z$ values on the rest-frame central
surface brightness is shown in Fig.~\ref{thickmu}. Thin disks
are present for $\mu_0(B) \geq 21{^{\rm mag}}/\sq\arcsec$, while among bright
objects ($\mu_0(B) < 21{^{\rm mag}}/\sq\arcsec$) we observe only thick disks 
with $h/z_0 \approx 2.2$. Taking into account that the
$B$-band surface brightness characterises the current rate
of star formation, one can speculate that enhanced star-formation
and enhanced thickness are the results of ongoing interactions
and mergings. The observed morphology of galaxies does not
contradict such a suggestion (Fig. 8).  

\subsection{Vertical scale for non-exponential galaxies}

We examined the vertical profiles for 15 non-exponential galaxies
at distances where the surface brightness drops by  1$^{\rm m}$ relative
to the nucleus. The mean observational value of the scaleheight 
obtained ($\langle h_z \rangle =0\farcs125$)
is the same as for exponential objects (Sect.3.3) with a  mean dispersion of 
individul values of 0\farcs03. Taking into account that the redshift 
and angular diameter distributions 
in both subsamples are approximately the same, one can 
conclude that non-exponential disks must have approximately
the same corrected value of the scale height ($h_z \sim$0.5 kpc).

\section{Discussion and conclusions}

Our sample of distant edge-on galaxies shows an unexpectedly large fraction 
of non-exponential disks. From this observation we infer that many 
sub-$L^*$ disks are still in the
formation process at $z \sim 1$, or the degree
of tidal perturbations is higher, which has
already been observed (e.g. Le F\'{e}vre et al. 2000). 
One of the possible mechanisms to
redistribute matter and to develop the exponential profile is the action of
a long-lived bar (e.g. Combes et al. 1990). 
The young galaxies of our sample may not have experienced 
enough bar torques for this process, either because of
very transient bars in very unstable disks, or because the
effects of bars are continuously perturbed by mergers and
accretion. The consequence is that a galaxy will
spend much less of its time in a barred state.
This conclusion is in good agreement with the general depletion of barred
spirals at $z > 0.7$ (Abraham et al. 1999, van den Bergh 2002).

We have taken into account
the effect of the PSF to demonstrate that the distant disks examined 
in this work are thicker by a factor of 1.5 if compared to  local disk sample.
This is a preliminary result based on a small number of sub-$L^*$ galaxies, 
and should be confirmed by larger statistical samples.

Since the star formation rate increases strongly with 
redshift (e.g. Madau et al. 1996, Flores et al. 1999, Thomson et al. 2001)
one could  speculate that the 
fraction of young stars formed in the midplane of disks
is expected to be much higher in the past, making the young disk
appear much thinner. 
Previous work on local samples has already 
concluded that tidal interactions between galaxies 
have a certain thickening effect of the order of 1.5-2
(Reshetnikov \& Combes 1997, Schwarzkopf \& Dettmar, 
2000). Since the frequency of galaxy interactions/mergers
increases strongly with redshift, the present 
result fits into this picture. If one assumes that 
all disks have undergone such interactions, either additional
cooling processes for the hot disks or a substantial growth
predominantly in the radial direction are required to explain the 
flatness of disks in the local Universe.    

However, other parameters should be taken into account 
in the interpretation. The distant disks are on absolute scales
smaller than the local sample disks. It is likely that 
the morphological types are later (less evolved) and the masses of
the galaxies smaller on average, as already found in
previous studies based on HST surveys (e.g. Abraham et al. 1996,
Glazebrook et al. 1998). The lower frequency of bars,
associated with the smaller bulge-to-disk ratio, suggests
that disks are more unstable and dynamically hotter 
(Abraham et al. 1999), which could explain the present 
thickening effect. In this context it is also noteworthy to mention
that the galaxies of our sample contain only a few bulges as can be judged
from the major axis profiles in Fig.~8. This can either be interpreted as
evidence for a later formation of bulges by secular processes (e.g. Combes 2001)
or that our sample is selected such that it represents early phases of
 late morphological types only.  

\acknowledgements{ VR acknowledges support by DAAD for his extended stay
at Ruhr-University Bochum and by Russian Federal Program ``Astronomy''
(40.022.1.1.1101). We thank Michael Pohlen for his help improving
the text. We would like to thank the anonymous referee whose detailed remarks 
have helped improve the paper.}

\section{Appendix}

We present here contour maps and major axis photometric profiles for
the sample of edge-on galaxies in the HDF (Table 1).

\begin{figure*}[h]
\centerline{\psfig{file=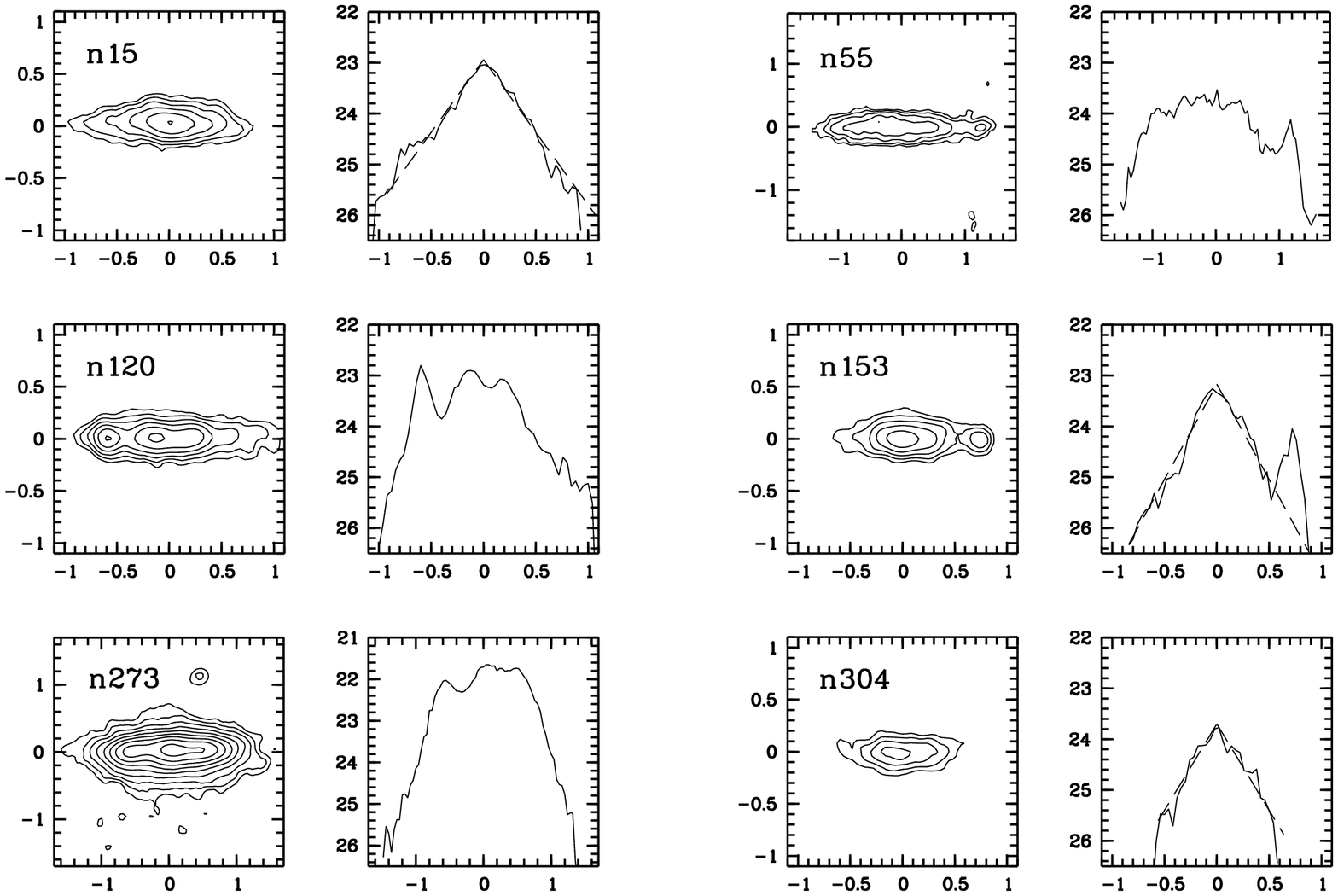,width=15.0cm,angle=-90,clip=}}
\label{cont1}
\end{figure*}

\begin{figure*}
\centerline{\psfig{file=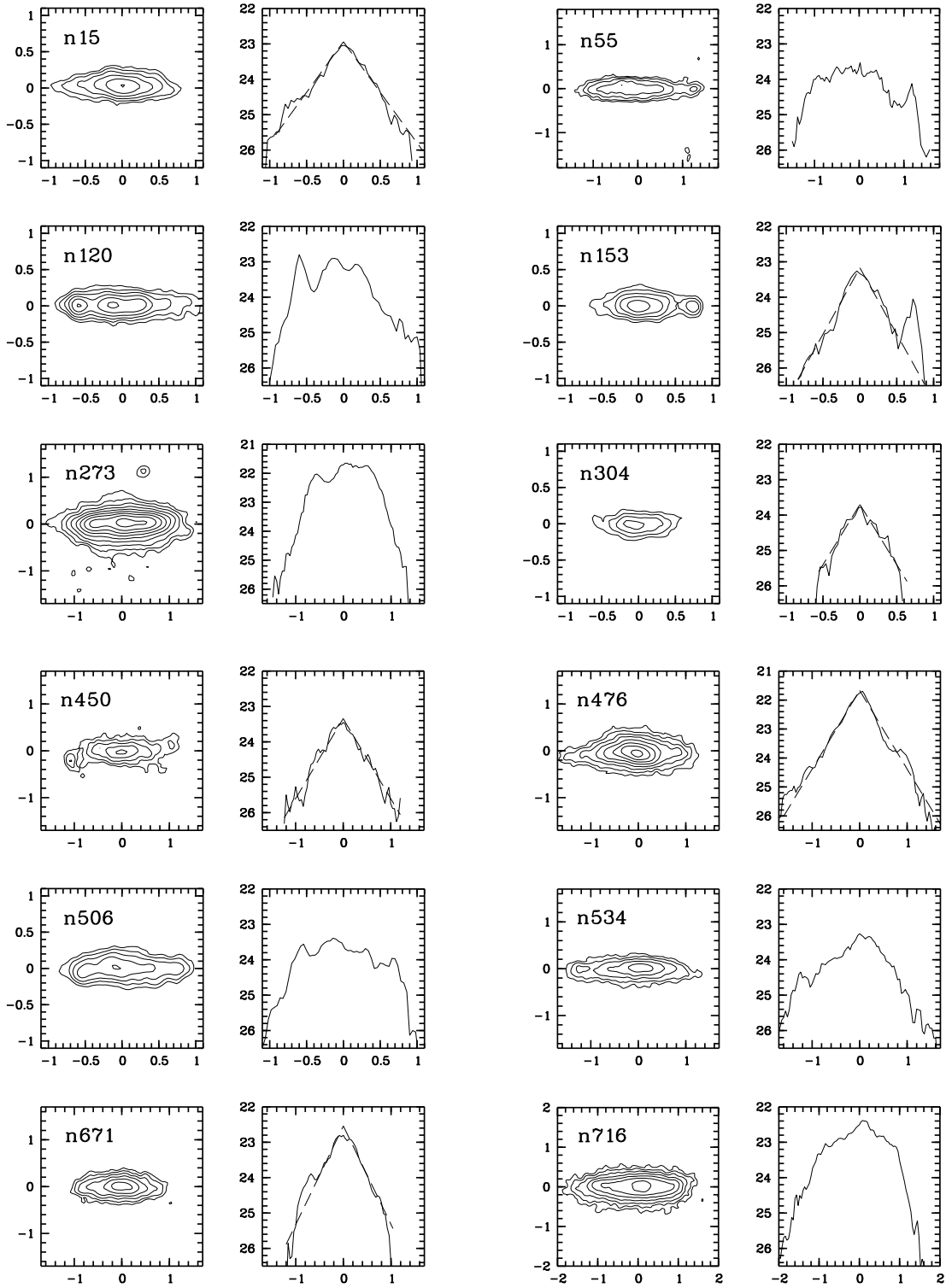,width=15.0cm,angle=-90,clip=}}
\caption{Contour plots (left) and major axis photometric profiles 
(right) of edge-on galaxies in the $I_{814}$ passband. The faintest 
contour corresponds to 2$\sigma$ limiting isophote of the HDF or
$\mu(I_{814})\approx25.75$, isophotes step is 0.$^m$5. 
Dashed lines -- symmetric exponential model for the disk.} 
\label{cont2}
\end{figure*}

\begin{figure*}
\centerline{\psfig{file=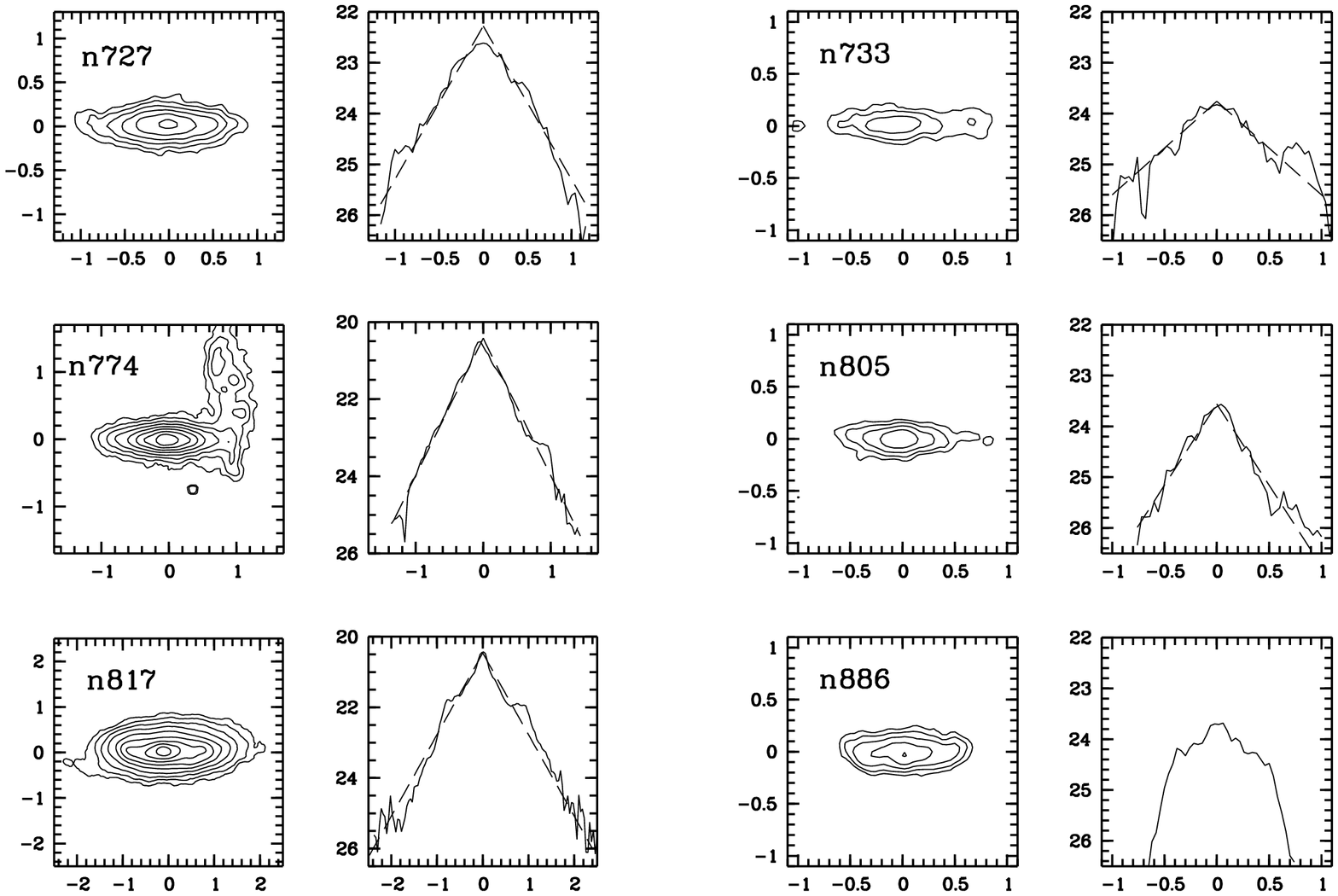,width=15.0cm,angle=-90,clip=}}
\label{cont3}
\end{figure*}
 \setcounter{figure}{7}
\begin{figure*}
\centerline{\psfig{file=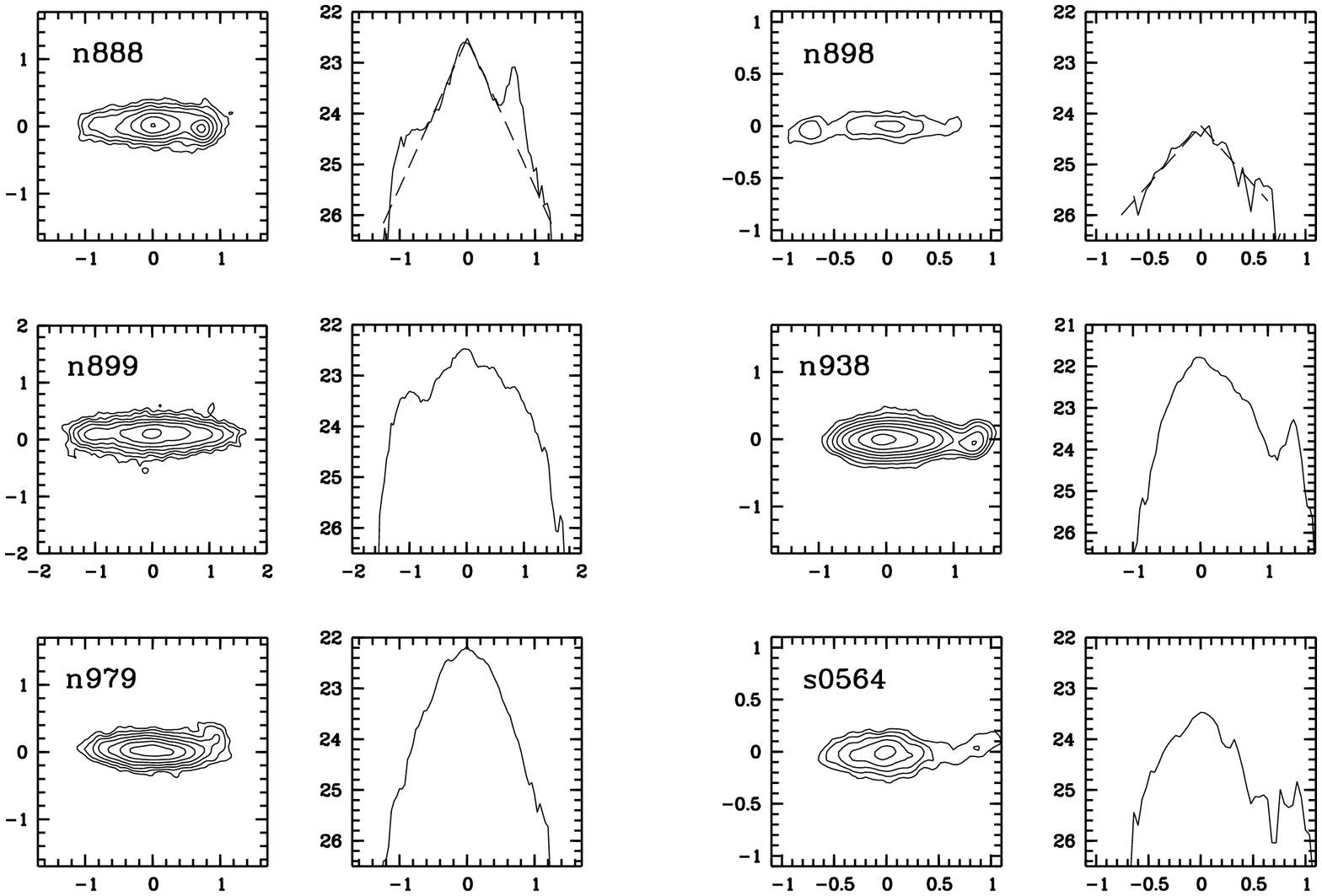,width=15.0cm,angle=-90,clip=}}
\caption[Fig. 8 cont.]{(cont.)}
\label{cont4}
\end{figure*}

\begin{figure*}
\centerline{\psfig{file=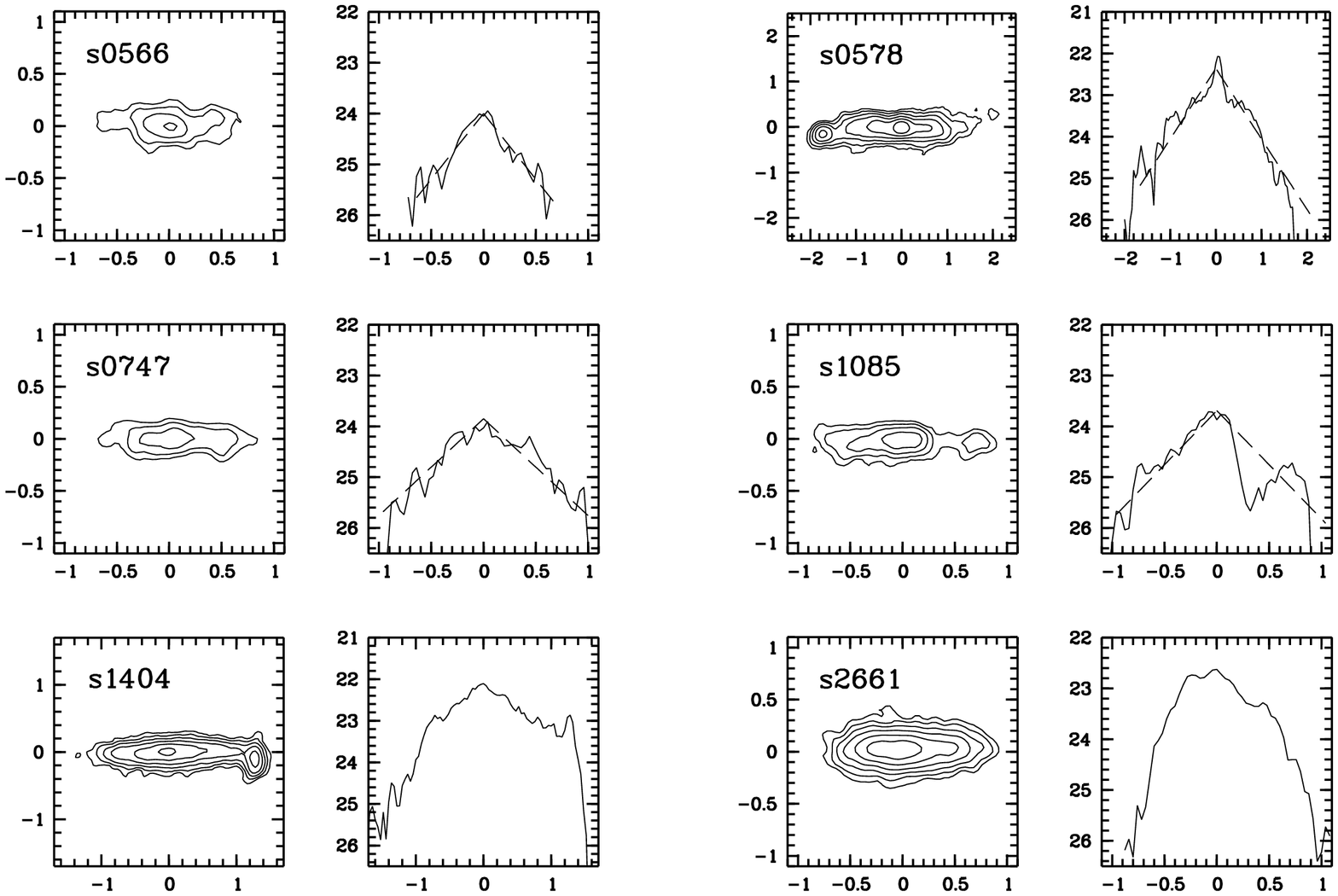,width=15.0cm,angle=-90,clip=}}
\label{cont5}
\end{figure*}
 \setcounter{figure}{7}
\begin{figure*}
\centerline{\psfig{file=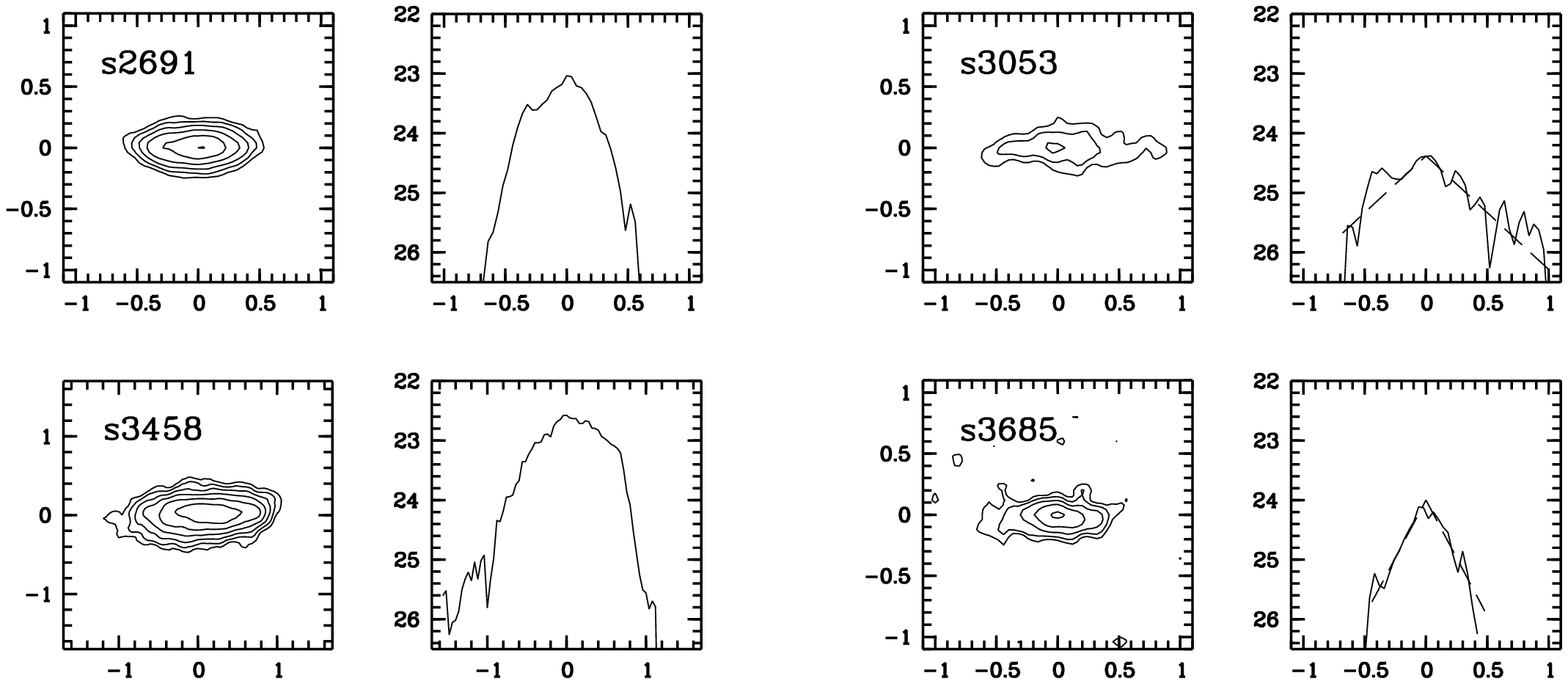,width=15.0cm,angle=-90,clip=}}
\caption{(cont.)}
\label{cont6}
\end{figure*}

\end{document}